\documentclass[aps,preprint,notitlepage]{revtex4-1}

\usepackage{graphicx}

\begin{document}

\title{Light guiding above the light line in arrays of dielectric nanospheres}
\author{Evgeny N. Bulgakov}
\author{Dmitrii N. Maksimov}
\affiliation{Kirensky Institute of Physics, 660036, Krasnoyarsk, Russia}

\begin{abstract}
We consider light propagation above the light line in arrays of spherical dielectric nanoparticles. It
is demonstrated numerically that quasi-bound leaky modes of the array can propagate both stationary waves
and light pulses to a distance of hundreds wavelengths at the frequencies close to the bound states in the radiation continuum.
A semi-analytical estimate  for decay rates of the guided waves is found to match the numerical data to a good accuracy.
\end{abstract}


\maketitle


The progress in subwavelength optics opens unprecedented opportunities for manipulating light on the nanoscale
\cite{Barnes2003,Novotny2012,Park2013}. Among those is the fabrication of subwavelength waveguides which may serve as
the key components for future integrated optics \cite{Quinten1998,Law2004,Pile2005,Skorobogatiy2012,Guo2014}. Since the seminal paper
by Quinten and co-authors \cite{Quinten1998} one of the mainstream ideas in the design of subwavelength waveguides
has been the implementation of various assembles of plasmonic nanoparticles such as considered in \cite{Pile2005,Brongersma2000, Maier2003,
 Weber2004, Alu06, Liu10, Rujting2011, Rasskazov2013} to name a few relevant references from the vast literature on the
subject. Seemingly less attention, though, has been payed to the arrays of dielectric nanonparticles \cite{Burin2004,Blaustein2007,Gozman2008,
Zhao2009,Du09,Du2011} although all-dielectric nanooptics  \cite{Krasnok2015, Savelev2015a} could be potentially advantageous against
nanoplasmonics due to, for instance, the opportunity to control the frequencies of electric and magnetic
Mie resonances by changing the geometry of high-index nanoparticles, and the absence of free carriers resulting in a high Q-factor. Arguably,
the arrays of dielectric nanoparticles provide one of the most promising subwavelength set-ups for efficient light
guiding \cite{Du09,Du2011,Savelev14} as well as more intricate effects such as resonant transmission of light \cite{Savelev2015},
and optical nanoantennas \cite{Krasnok2012}.

So far the major theoretical tool for analyzing the infinite arrays of spherical dielectric nanoparticles has been the coupled-dipole approximation
\cite{Merchiers07, Evlyukhin10, Wheeler2010a}. In that approximation guided waves in arrays of magnetodielectric spheres were
first considered by Shore and Yaghjian \cite{Shore2004, Shore05} who derived the dispersion relation and computed the dispersion curves for
dipolar waves. Recently a more tractable form of the dispersion equations was
presented by the same authors \cite{Shore2012a} with the use of the polilogarithmic functions. The dipolar waves in arrays
of Si dielectric nanospheres were thoroughly analyzed in \cite{Savelev14}. In particular, it was shown that only two lowest guided
modes could be fairly described by the dipole approximation which breaks down as the frequency approaches the first quadruple Mie resonance.
This limits the application of the dipolar dispersion diagrams to realistic waveguides assembled
of dielectric nanoparticles. As an alternative to the dipole approximation a "semiclassical" approach based on the coupling of the whispering
gallery modes of individual spheres could be employed to recover the array band structure \cite{Deych2005,Deych2006} if the wavelength
is much smaller than the diameter of the spheres. The general case, however, requires a full-wave Mie scattering approach to account
for all possible multipole resonances \cite{Blaustein2007} involving a very complicated multiscattering picture which
mathematically manifests itself in infinite multipole sums. Luckily, such an approach was recently developed by Linton,
Zalipaev, and Thompson who managed to obtain a multipole dispersion relation in a closed form suitable for
numerical computations  \cite{Linton13}. The above approach was used for analyzing the spectra of dielectric arrays above the line of light
in ref. \cite{Bulgakov15}. It was demonstrated that under variation of some parameter such as, for example, the radius of the spheres
the leaky modes dominating the spectrum can acquire an infinite life-time. In other words, the array can support bound states in the radiation
continuum (BSCs) \cite{Ndangali2010,Chia_Wei_Hsu13,Bulgakov2014,Monticone2014,Gao2016}.
In this letter, we will address the ability of the BSCs to propagate light along the array primarily motivated by finding new opportunities for designing subwavelength waveguides.

\begin{figure}
\includegraphics[width=0.5\textwidth,trim={0cm 0cm 0 0},clip]{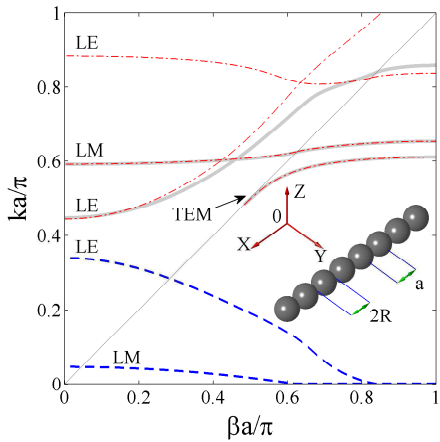}
\caption{(Color online) Dispersion diagram of an infinite array of dielectric nanospheres of radius $R$ with dielectric constant $\epsilon=15$, $R/a=0.4$.
The array centerline is aligned  with the $x$-axis as shown in the south-east corner of the plot.
The real parts of dipolar solutions are shown by dash-dot red lines. The thick gray lines are the real parts of the full-wave solutions;
negative imaginary parts $-\Im(k)$ of the full-wave solutions are shown by blue dashed lines. The thin gray line is the line of light.}\label{Fig1}
\end{figure}

Let us first obtain the dispersion diagram of an array of dielectric nanoparticles. The dispersion curves are computed
by solving the dispersion equations $f_{d,m}(k,\beta)=0$, where $k$ is the vacuum wave number $k=\omega/c$, and $\beta$ is the
Bloch wave number, while the subscripts $d,m$ designate
either dipole \cite{Shore2012a,Savelev14}, or multipole \cite{Linton13} dispersion relations. For brevity we do not present
the exact dispersion relations $f_{d,m}(k,\beta)=0$. A mathematically inquisitive  reader is referred to the above cited papers
to examine the rather cumbersome expressions for $f_{d,m}(k,\beta)$. Here we assume that the array consists of spherical noanoparticles of radius $R$ with dielectric
 constant $\epsilon=15$ (Si) in vacuum. The centers
of the nanoparticles are separated by distance $a$. It is worth mentioning that at a given dielectric constant the dispersion is
only dependent on a single dimensionless quantity $R/a$. This allowed to scale the model for a microwave experiment  \cite{Savelev14}.
There are three types of dipolar solutions \cite{Savelev14}, namely; longitudinal magnetic (LM), longitudinal electric (LE), and transverse
electromagnetic (TEM) waves. In Fig. \ref{Fig1} we plot the lowest frequency modes of each type in comparison against the multipole solution
\cite{Linton13}. In all cases if the $k-\beta$ curve is above the light line $k=\beta$ the vacuum wave number becomes complex valued.
The imaginary part of $k$ is linked to the mode life-time through the following formula
\begin{equation}\label{tau}
\tau=-[c \Im(k)]^{-1}.
\end{equation}
Two approaches are possible for description of the leaky modes; complex frequency $\omega$ \cite{Tikhodeev2002,Bykov2013}, or complex
Bloch number $\beta$ \cite{Shore2012b, Savelev14}. In the latter case the inverse of the imaginary part of $\beta$
is the penetration depth into the array $L_{\tau}=[\Im(\beta)]^{-1}$. The quantities $\tau$ and $L_{\tau}$ are, in fact, proportional
\begin{equation}\label{penetration_depth}
L_{\tau}=v \tau,
\end{equation}
where $v$ is the group velocity $v=d\Re(\omega)/d\beta$. Here, we do not present the imaginary part of $\beta$ mentioning in passing
that the penetration depths for dipolar waves were analyzed in refs. \cite{Shore2012b, Savelev14}. What is important the numerical data
available so far \cite{Shore2012b, Savelev14, Bulgakov15} indicate that all dipolar leaky modes are relatively short-lived, in particular, no
dipolar BSCs were found in ref. \cite{Bulgakov15}. In compliance with ref. \cite{Savelev14} Fig. \ref{Fig1} demonstrates that only two lowest
eigenmodes are fairly described by the dipole approximation. Thus, for obtaining valid results for guided modes
at $ka/\pi\approx1$ one has to resort to the full-wave formalism of ref. \cite{Linton13}.
\begin{figure}[t]
\begin{center}
\includegraphics[width=0.6\textwidth, height=0.6\textwidth,trim={4cm 1.5cm 0cm 0cm},clip]{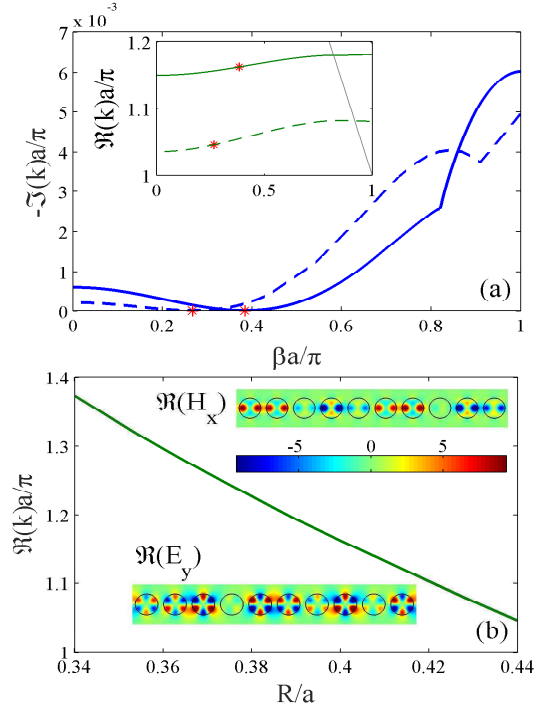}
\caption{(Color online) (a) Quasi-guided modes above the light line; $R/a=0.4$-solid line, $R/a=0.44$-dash line, $\epsilon=15$. Imaginary part of $k$ - the main plot, real
part - the inset. The positions of the BSCs are shown by red stars.  The imaginary parts are non-smooth
as the real parts cross the boundary of the second radiation continuum $ka=2\pi-\beta{a}$ shown by thin grey line.
(b) Bloch BSC $\beta\neq0$ vacuum wave number $k$ vs. $R/a$, $\epsilon=15$. The insets show the real parts of the $y$-component of electric vector $E_y$ and the $x$-component
of magnetic vector $H_x$ in $x0y$-plane for the BSC at $R/a=0.4$.}\label{Fig2}
\end{center}
\end{figure}

Now, let us consider the multipolar quasi-guided modes within the first radiation continuum \cite{Bulgakov15}. The dispersion curves
for a leaky mode for two different ratios $R/a$ are plotted in Fig. \ref{Fig2} (a).
One can see that in contrast to the dipolar waves
in Fig. \ref{Fig1} now the solutions could be long-lived with the life-time Eq. (\ref{tau}) growing up to infinity at the BSC points.
It should pointed out that for both $R/a$ of all leaky modes of the array we plot only one which has a Bloch BSC point $\Im(k)=0$ at $\beta\neq0$.
As shown in Fig. \ref{Fig2} (b) the BSC exists in a wide range of parameter $R/a$.
The magnetic and electric vectors could be found in terms of Mie coefficients $a_n^m, b_n^m$. For instance, outside the spheres
one has for the electric vector ${\bf E}({\bf r})$ \cite{Linton13}
\begin{equation}\label{expansion}
{\bf E}({\bf r})=\sum^{\infty}_{j=-\infty}e^{iaj \beta}\sum^{\infty}_{n=m^{*}}
\left[a_n^m{\bf M}^{m}_{n}({\bf r}-{\bf r}_j)+b_n^m{\bf N}^{m}_{n}({\bf r}-{\bf r}_j)\right],
\end{equation}
where $j$ the number of the particle in the array, m - azimuthal number, $m^*=max(1,m)$,
and ${\bf N}^{m}_{n}({\bf r}),{\bf M}^{m}_{n}({\bf r})$ are spherical
vector harmonics \cite{Stratton41}. Only $m=0$ Bloch BSCs were found in ref. \cite{Bulgakov15}. Our numerics indicate that for BSCs in
Fig. \ref{Fig2} the dominating term in the expansions (\ref{expansion})
corresponds to coefficient $a_3^0$ . In the insets in Fig. \ref{Fig2} (b) we plot the components of the
electric and magnetic vectors of the BSC solution. One can see that the electromagnetic field is localized in the vicinity
of the array.

The amplitude of a wave
propagating along the array attenuates exponentially according to a simple formula
\begin{equation}\label{attenuation}
F(x)=e^{-x/L_{\tau}},
\end{equation}
where $x=ja$ is the distance.
In the vicinity of a BSC the $\omega-\beta$ dependance could be approximated
as \begin{equation}\label{frequency}
\omega-\omega_{0} = v_0(\beta-\beta_{0}),
\end{equation}
where $\omega_{0}, \beta_{0}, v_0$ are the BSC eigenfrequency, Bloch number, and group velocity, correspondingly.
For the imaginary part of the vacuum wave number the leading term is, however, quadratic
\begin{figure}[t]
\includegraphics[width=0.35\textwidth]{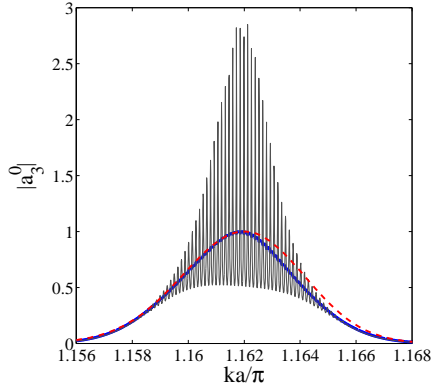}
\caption{(Color online) Absolute value of the leading coefficient $a_3^{0}$ for the last nanoparticle
vs. wave number $k$ for waves penetrating into the array of $400$ nanoparticles; $\epsilon=15$, $R/a=0.4$, $v_0=0.054c, \alpha=0.005a$.
The averaged data are plotted by thick blue line against the analytical result
Eq. (\ref{attenuation_finale}) shown by dashed red line.}\label{Fig3}
\end{figure}
\begin{equation}\label{imaginary_part}
-\Im\{k\}=\alpha(\beta-\beta_0)^2.
\end{equation}
Combining Eqs. (\ref{tau},\ref{penetration_depth},\ref{attenuation},\ref{frequency},\ref{imaginary_part}) one
obtains
\begin{equation}\label{attenuation_finale}
F(x)=\exp \left[-\frac{\alpha x c}{v_0^3}(\omega-\omega_0)^2\right].
\end{equation}
Thus, for the width of the transparency window in the frequency domain we have
\begin{equation}
\Delta(x)=\sqrt{\frac{v_0^3}{\alpha c}}\frac{1}{\sqrt{x}}.
\end{equation}

Using a full-wave multiscattering method \cite{Blaustein2007} we simulated wave propagation in a finite array of $400$ nanoparticles.
In our numerical experiment a linearly polarized Gaussian beam \cite{Carrasco2006} with the Rayleigh range $z_0=5a$ was focused on the first
nanoparticle in the array. The wave vector
of the beam was directed along the $y$-axis perpendicular to the array (see. Fig. \ref{Fig1}), and the magnetic vector aligned with
the array axis. In Fig. \ref{Fig3} we plot the
the leading Mie coefficient $a_{3}^{0}$ for the last nanoparticle in the array. The result shows a pronounced resonant behavior due
to formation of standing waves as a consequence of the finiteness of the array. The distance between the resonances
$\Delta \omega$ could be assessed as $\Delta \omega \approx \pi v_0/(a N)$ where $N$ is the number of particles in the array.
The resonant features could be averaged out by integration over small frequency intervals larger than $\Delta \omega$.
The result is shown in Fig. \ref{Fig3} in comparison against Eq. (\ref{attenuation_finale}). One can see that Eq. (\ref{attenuation_finale}) matches the numerical data to a good accuracy.

\begin{figure}[t]
\includegraphics[width=0.40\textwidth]{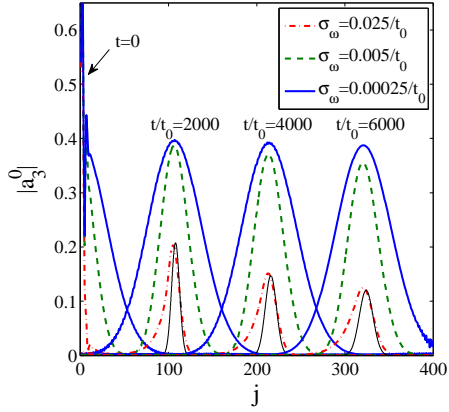}
\caption{(Color online) Pulse propagation along the array of $400$ nanoparticles. Absolute value of the leading coefficient $a_3^{0}$ vs.
the number nanoparticle number; $k_0=\pi 1.162/a$, $a=309$nm $t_0=1.77$fs. Thin black line shows analytical result Eq. (\ref{pulse}) for
the pulse with $\sigma_{\omega}=0.025/a$. The other parameters are the same
as in Fig. \ref{Fig3}.}\label{Fig4}
\end{figure}
Finally, pulse propagation along the array was considered. The above set-up was retained with a continuous superposition
of Gaussian beams forming a Gaussian light pulse of width $\sigma_{\omega}$ in the frequency domain.
The central wave number of the pulse was adjusted to the BSC wave number $k_0=\pi 1.162/a$. So far we intentionally presented the results
as dimensionless quantities $ka,{\beta}a$ to underline the scaling properties of the system under consideration.
Now we specify the vacuum wavelength for determining the length and time scales typical for
a realistic optical experiment. Let us choose $\lambda=532$nm which is the vacuum wavelength of the second harmonics Nd:YAG laser.
This corresponds to $a=309$nm and $R=124$nm. Let us also define the characteristic time
$t_0=a/c=1.77$fs. At the moment $t=0$ a light pulse was injected into the left end of the array. The response of the system was again
recorded in terms of the Mie coefficients. In Fig. \ref{Fig4} we plot four snapshots of the leading Mie coefficient $a_{3}^{0}$ against
the distance along the array for three different initial pulse widths $\sigma_{\omega}$. One can clearly see in Fig. \ref{Fig4} that the
pulse propagating along the array tends to spread as the harmonics distant in the $\omega$-space from the BSC frequency decay into
the radiation continuum. The pulse profile $f(x,t)$ could be found by Fourier-transforming the initial Gaussian pulse to the real space
\begin{equation}\label{pulse}
f(x,t)=\frac{1}{\sigma(t)}e^{-\frac{(x-v_0t)^2}{\sigma^2(t)}}e^{i(\beta_0x-\omega_0t)}
\end{equation}
with
\begin{equation}\label{dispersion}
\sigma^2(t)=4a^2\left[\left(\frac{v_0}{a\sigma_{\omega}}\right)^2+\frac{c{\alpha}}{a^2}t\right].
\end{equation}
Analyzing Eq. (\ref{dispersion}) for a given detection distance $L=v_0t$ one can identify two possible
regimes for the pulse propagation. In the "overdamped" regime the second term
dominates on the left hand side of Eq. (\ref{dispersion}) resulting in a noticeable spreading of the pulse in the real space. If, however, the first term
dominates the pulse retains its profile during propagation time. Thus, tuning $\sigma_{\omega}$ one can achieve a propagation distance $L=400a \approx 200\lambda$
without a significant distortion of the pulse profile ($\sigma_{\omega}=0.0025/t_0$ in Fig. \ref{Fig4}).
One finds from Eq. (\ref{dispersion}) that the pulse doubles its width after travelling to the distance
\begin{equation}\label{distance}
L=\frac{3a^2}{\alpha}\left(\frac{v_0}{c}\right)^3\left(\frac{c}{a\sigma_{\omega}}\right)^2.
\end{equation}

In summary, we demonstrated the effect of light guiding above the light line in arrays of spherical dielectric lossless
nanoparticles. For the guiding of light we employed leaky modes residing in the radiation continuum. It was found
that long-lived leaky modes are associated with bound states in the radiation continuum which are supported by the arrays in a broad range
of parameters (see Fig. \ref{Fig2}). It should be pointed out that the reported mutiplolar solutions, though still in the subwavelength range,
have a higher $R/\lambda$ ratio than the dipolar
solutions in ref. \cite{Savelev14}. This, on the other hand, relaxes the condition $\lambda>2a$ for guided waves in arrays and gratings
\cite{Blaustein2007}.
Besides the fundamental aspects, the benefits of employing leaky modes could
be the opportunity to guide light harvested from free waves propagating in the ambient medium \cite{Bulgakov15, Bulgakov2014}
and potential capacity to propagate multiple frequencies of light both below and above the radiation continuum in the same
subwavelength structure.

This work has received financial support from RFBR through grant 16-02-00314.
We acknowledge discussions with A.F. Sadreev, A.S. Aleksandrovsky, and A.M. Vyunishev.

\bibliography{Subwavelength_waveguides}

\end{document}